\documentclass[12pt]{article} %{slides}

\usepackage{amsmath,amssymb,amsfonts,pstricks,pst-plot,pst-node,a4wide}
\usepackage{layout,array}
\usepackage[T1]{fontenc}
\newcommand{\R}{\ensuremath{\mathbb R}}

\newtheorem{theorem}{Theorem}
\def\bar{\overline}
\def\Hl{\xi_\ell}
\def\Hn{\xi_n}
\def\Hm{\xi_m}
\def\Hmb{\xi_{\bar m}}
\def\mb{\bar m}
\def\o{{o}}
\def\i{{\iota}}
\def\sfdual{{}^{\raise2pt\hbox{$\scriptstyle{-}$}}}

\def\GHPL{\operatorname{\mathcal L}}

\def\thorn{\operatorname{\text{\th}}}
\def\eth{\operatorname{\text{\dh}}}

\begin{document}
\begin{center} % TITLE
{\large\bf Conformal vectors in general space-times}
\end{center}

\begin{narrower}

\medskip
\noindent {\bf John D.~Steele}

\smallskip
\noindent {\small School of Mathematics and Statistics, University of New South
Wales,
Sydney, NSW 2052, Australia.

\smallskip
\noindent  email: j.steele@unsw{.}edu{.}au}

\medskip
\noindent {\bf Abstract} 

\noindent 
In an earlier paper [14] the author wrote the homothetic equations for vacuum
solutions in a first order formalism allowing for arbitrary alignment of the
dyad. This paper generalises that method to conformal vectors in non-vacuum
spaces. The method is applied to metrics admitting a
three parameter motion group on non-null orbits.

\medskip
\noindent PACS number 0420J

\medskip
\noindent Mathematics Subject Classification: 83C15, 83C20

\end{narrower}

\section{Introduction and Notation}

In [14] I gave the homothetic Killing equations in vacuum, written out in a
first order form without the assumption that the spinor dyad used was aligned to
either the symmetry or the curvature in any way, and indeed allowing for the
dyad to be non-normalised.
In this paper I will generalise to conformal vectors in non-vacuum spaces-times. 
Equations for this case have been given in [9,11] in a particularly specialised
notation, although the former did not use a first order formalism.

A conformal vector $\xi^a$ by definition satisfies the equation
\begin{equation}\xi_{a;b}= F_{ab} + \psi g_{ab}\label{eq:Killing}.\end{equation}
Here $\psi$, the {\bf divergence}, is an arbitrary scalar, and $F_{ab}$ will be
called the {\bf conformal bivector}. If $\psi$ is constant we have a homothetic
vector, if $\psi_{,a}$ is covariantly constant we have a {\bf special conformal
vector} [16].

Let $\{\o^A,\i^A\}$ be a spinor dyad, with
$\o_A\i^A=\chi$. A complex null tetrad is related to this dyad in the standard 
way:
$$\ell^a=\o^A\o^{A'};\quad n^a=\i^A\i^{A'};\quad m^a=\o^A\i^{A'};\quad
\mb^a=\i^A\o^{A'},$$
([15], (4.5.19)), and $\ell_an^a = -m_a\mb^a = \chi\bar\chi$.
As in [14], we define components of the conformal:
\begin{equation}
\xi_a =\Hn\ell_a + \Hl n_a - \Hmb m_a - \Hm \mb_a,
\end{equation} 
with $\{\ell^a,n^a,m^a,\mb^a\}$ a
Newman-Penrose tetrad. Thus, for example, $\chi\bar\chi\Hl=\xi_a\ell^a$.

For the conformal bivector
$F_{ab}$ we define its anti-self dual by
\begin{align}
\sfdual{F}_{ab}& =\frac{1}{2}\left(F_{ab}+i{}F^*_{ab}\right)
\end{align}
and then
\begin{equation}
\sfdual{F}_{ab} =(\chi\bar\chi)^{-1}\left( 2\phi_{00}\,\ell_{[a}m_{b]}
 + 2\phi_{01}\,(\ell_{[a}n_{b]} - m_{[a}\mb_{b]})
 - 2\phi_{11}\,n_{[a}\mb_{b]}\right),
\end{equation} 
where
\begin{align}
\phi_{11} & = (\chi\bar\chi)^{-1}F_{ab}\ell^am^b \\
\phi_{01} & = \frac{1}{2}(\chi\bar\chi)^{-1}\left(
 F_{ab}n^a\ell^b - F_{ab}\mb^am^b\right) \\
\phi_{00} & = (\chi\bar\chi)^{-1}F_{ab}\mb^a n^b
\end{align}

Most of the equations in this paper will be given using the compacted
GHP-formalism, see [5,15,16]. In
this formalism, we simplify notation by concentrating on those spin coefficients of
good weight, that
is, those that transform homogeneously under a spin-boost transformation of the
dyad: if
$$ \o^A\mapsto \lambda \o^A \qquad \i^A\mapsto \mu\i^A$$
a {\bf weighted} quantity $\eta$ of {\bf type} $\{r',r;t',t\}$ undergoes a 
transformation
$$  \eta \mapsto \lambda^{r'}\bar\lambda^{t'} \mu^{r}\bar\mu^{t}\eta.$$
These weights will be referred to as the {\bf Penrose-Rindler (PR)} weights, to
distinguish them from the more familiar GHP-weights $(p,q)$ in e.g.~[5,16],
where $p=r'-r$ and $q=t'-t$. A  second advantage of this notation is that it is
indifferent to the scaling of the tetrad.

\section{General Equations}

The conformal equations themselves, (\ref{eq:Killing}), are unaffected by the
curvature or the fact that $\psi$ is not constant and so are the same as in [14]:
\newcounter{compkilleq}
\stepcounter{equation}
\setcounter{compkilleq}{\value{equation}}
\begin{align}
\thorn\Hl & = -\bar\kappa\Hm - \kappa\Hmb;   \tag{\thecompkilleq$a$}\\
\thorn'\Hl & =  -\bar\tau\Hm -\tau\Hmb 
     -(\phi_{01}+\bar\phi_{01})+\psi; \tag{\thecompkilleq$b$} \\
\eth\Hl &= -\bar\rho\Hm-\sigma\Hmb  +\phi_{11}; \tag{\thecompkilleq $c$}\\
\thorn\Hn &= -\tau'\Hm-\bar\tau'\Hmb 
     +(\phi_{01}+\bar\phi_{01})+\psi; \tag{\thecompkilleq $d$} \\
\thorn'\Hn &=  - \kappa'\Hm -\bar\kappa'\Hmb ;  
   \tag{\thecompkilleq$e$}\\
\eth\Hn &=  -\rho'\Hm-\bar\sigma'\Hmb    -\bar\phi_{00};
   \tag{\thecompkilleq$f$}\\
\thorn\Hm & = -\bar\tau'\Hl - \kappa\Hn  -\phi_{11};\tag{\thecompkilleq$g$}\\
\thorn'\Hm & = -\bar\kappa'\Hl - \tau\Hn +\bar\phi_{00}; 
   \tag{\thecompkilleq$h$}\\
\eth\Hm &= -\bar\sigma'\Hl -\sigma\Hn  ;   
    \tag{\thecompkilleq$i$}\\
\eth'\Hm & = -\bar\rho'\Hl-\rho\Hn 
    +(\phi_{01}-\bar\phi_{01})- \psi; \tag{\thecompkilleq$j$}
\end{align}

The spin-boost weights $(r',r,t',t)$ of the components of $\xi^a$ and $F_{ab}$  are
given in Table~\ref{tb:one} (correcting a minor typo in [14]):

\begin{table}[h]
$$\vbox{
\tabskip=0pt\offinterlineskip
\halign %to \hsize
{\strut#&\vrule # \tabskip=.75em plus 2em minus 0.5em
  &\hfil $#$\hfil&\vrule# &\hfil $#$\hfil&\vrule# 
  &\hfil $#$\hfil&\vrule# &\hfil $#$\hfil&\vrule# 
  &\hfil $#$\hfil&\vrule# &\hfil $#$\hfil&\vrule# 
%  &\hfil #\hfil&\vrule# &\hfil #\hfil&\vrule# 
  &\hfil $#$\hfil&\vrule# &\hfil $#$\hfil&\vrule# \tabskip=0pt\cr
\noalign{\hrule}
\omit & height 3pt && height 3pt &&  height 3pt && %height 3pt && height 3pt &&
 height 3pt && height 3pt  && height 3pt && height 3pt && height 3pt 
&& height 3pt\cr
&&   && \Hl && \Hn && \Hm && \Hmb && \phi_{00} && \phi_{01} && \phi_{11} &\cr
\noalign{\hrule}
\omit & height 3pt && height 3pt &&  height 3pt && %height 3pt && height 3pt &&
 height 3pt && height 3pt  && height 3pt && height 3pt && height 3pt 
&& height 3pt\cr
&& r' && 0 && -1 && 1 && 0 && -1 && 0 && 1&\cr
\noalign{\hrule}
\omit & height 3pt && height 3pt &&  height 3pt && %height 3pt && height 3pt &&
 height 3pt && height 3pt  && height 3pt && height 3pt && height 3pt 
&& height 3pt\cr
&& r && -1 && 0 && -1 && 0 && 1 && 0 && -1&\cr
\noalign{\hrule}
 \omit & height 3pt && height 3pt &&  height 3pt && %height 3pt && height 3pt &&
  height 3pt && height 3pt  && height 3pt && height 3pt && height 3pt 
 && height 3pt\cr
 && t' && 0 && -1 && -1 && 0 && 0 && 0 && 0&\cr
 \noalign{\hrule}
\omit & height 3pt && height 3pt &&  height 3pt && %height 3pt && height 3pt &&
  height 3pt && height 3pt  && height 3pt && height 3pt && height 3pt 
 && height 3pt\cr
 && t && -1 && 0 && 0 && -1 && 0 && 0 && 0&\cr
 \noalign{\hrule}
}}$$
%\endinsert
\caption{weights of components}\label{tb:one}
\end{table}

The Ricci identity for $\xi^a$ implies $F_{cd;b}=R_{abcd}\xi^a-2\psi_{,[c}g_{d]b}$, 
from which
the algebraic Bianchi identities lead  to equations for the derivatives of the
$\phi_{ij}$.  
The anti-self-dual of this equation takes the
spinor form
\begin{align}
\nabla_{CC'}\phi_{AB}  = \left(\Psi_{ABDC}\epsilon_{D'C'} +
\Phi_{ABD'C'}\epsilon_{DC}\right)\xi^{DD'} &-
\Lambda\left(\epsilon_{BC}\xi_{AC'}+\epsilon_{AC}\xi_{BC'}\right)+{} \notag\\ 
&\frac12\left(\epsilon_{AC}\psi_{,BC'}+\epsilon_{BC}\psi_{,AC'}\right)
\label{eq:spinorintcond}
\end{align}
Here $\Psi_{ABCD}$ is the (totally symmetric) Weyl spinor, 
$\Phi_{ABA'B'}$ the Ricci spinor and $24\Lambda=R$, the Ricci scalar (see [15]).

The components of the Weyl and Ricci spinors are given in [15] (4.11.6) and
(4.11.8) respectively, and then resolving equation (\ref{eq:spinorintcond}) we
get the (first) integrability conditions
\newcounter{intcons}
\stepcounter{equation}
\setcounter{intcons}{\value{equation}}
\begin{align}
\Psi_1\Hl-\Psi_0\Hmb +\Phi_{01}\Hl -\Phi_{00}\Hm & 
  =2\kappa\phi_{01} %+(\eps-\gamma')\phi_{11}-D\phi_{11};
  -\thorn\phi_{11};
  \tag{\theintcons$a$}\\[1pt]
\Psi_1\Hm-\Psi_0\Hn +\Phi_{02}\Hl -\Phi_{01}\Hm &  
  = 2\sigma\phi_{01} %+(\beta-\alpha')\phi_{11}-\delta\phi_{11};
 -\eth\phi_{11};
  \tag{\theintcons$b$}\\[1pt]
\Psi_2\Hl-\Psi_1\Hmb +\Phi_{01}\Hmb -\Phi_{00}\Hn +2\Pi\Hl & 
  = 2\rho\phi_{01} %(+\alpha-\beta')\phi_{11}-\delta'\phi_{11};
  -\eth'\phi_{11} +\thorn\psi;
  \tag{\theintcons$c$}\\[1pt]
\Psi_2\Hm-\Psi_1\Hn +\Phi_{02}\Hmb -\Phi_{01}\Hn +2\Pi\Hm &
 = 2\tau\phi_{01} % +(\gamma-\eps')\phi_{11}-D'\phi_{11};
  -\thorn'\phi_{11} +\eth\psi;
  \tag{\theintcons$d$}\\[1pt]
\Psi_3\Hl -\Psi_2\Hmb +\Phi_{21}\Hl -\Phi_{20}\Hm -2\Pi\Hmb &
 = 2\tau'\phi_{01} %+(\gamma'-\eps)\phi_{00}-D\phi_{00};
   -\thorn\phi_{00}-\eth'\psi;
\tag{\theintcons$e$}\\[1pt]
\Psi_3\Hm-\Psi_2\Hn +\Phi_{22}\Hl -\Phi_{21}\Hm -2\Pi\Hn & 
 = 2\rho'\phi_{01} %+(\alpha'-\beta)\phi_{00}-\delta\phi_{00};
   -\eth\phi_{00} - \thorn'\psi;
 \tag{\theintcons$f$}\\[1pt]
\Psi_4\Hl -\Psi_3\Hmb +\Phi_{21}\Hmb -\Phi_{20}\Hn & 
 = 2\sigma'\phi_{01} %(+\beta'-\alpha)\phi_{00}-\delta'\phi_{00};
  -\eth'\phi_{00};
 \tag{\theintcons$g$}\\[1pt]
\Psi_4\Hm -\Psi_3\Hn +\Phi_{22}\Hmb -\Phi_{21}\Hn &  
 = 2\kappa'\phi_{01} %+(\eps'-\gamma)\phi_{00}-D'\phi_{00};
   -\thorn'\phi_{00}
\tag{\theintcons$h$}\\[1pt]
\Psi_2\Hl-\Psi_1\Hmb +\Phi_{11}\Hl -\Phi_{10}\Hm -\Pi\Hl & 
  = \thorn\phi_{01}-\tau'\phi_{11}-\kappa\phi_{00}-{\textstyle\frac12}\thorn\psi;
  \tag{\theintcons$i$}\\[1pt]
\Psi_2\Hm-\Psi_1\Hn +\Phi_{12}\Hl -\Phi_{11}\Hm -\Pi\Hm &
 = \eth\phi_{01}-\rho'\phi_{11}-\sigma\phi_{00}-{\textstyle\frac12}\eth\psi;
   \tag{\theintcons$j$}\\[1pt]
\Psi_3\Hl -\Psi_2\Hmb +\Phi_{11}\Hmb -\Phi_{10}\Hn +\Pi\Hmb &
 = \eth'\phi_{01}-\rho\phi_{00}-\sigma'\phi_{11}+{\textstyle\frac12}\eth'\psi;
   \tag{\theintcons$k$}\\[1pt]
\Psi_3\Hm-\Psi_2\Hn +\Phi_{12}\Hmb -\Phi_{11}\Hn +\Pi\Hn & 
 = \thorn'\phi_{01}-\tau\phi_{00}-\kappa'\phi_{11}+{\textstyle\frac12}\thorn'\psi.
 \tag{\theintcons$l$}
\end{align}
where $\Pi=\chi\bar\chi\Lambda$. These equations, which can also be obtained from
applying the commutators to equations~(\thecompkilleq), are equivalent to 
equations (20)--(22) in [11].

Note that there are four pairs of equations with the same Weyl curvature terms
($c$/$i$; $d$/$j$; $e$/$k$ and $f$/$l$).
We can eliminate the Weyl curvature terms between these pairs to give equations
equivalent to (23) in [11]:
\newcounter{maxintcons}
\stepcounter{equation}
\setcounter{maxintcons}{\value{equation}}
\begin{align}
\thorn\phi_{01} +\eth'\phi_{11} & - \kappa\phi_{00}  -2\rho \phi_{01}
-\tau'\phi_{11}-{\textstyle\frac32}\thorn\psi\notag\\
 &= \left(\Phi_{11}-3\Pi\right)\Hl +\Phi_{00}\Hn-\Phi_{10}\Hm-\Phi_{01}\Hmb
\tag{\themaxintcons$a$}\\[1pt]
\eth\phi_{01} +\thorn'\phi_{11} & - \sigma\phi_{00} -
  2\tau\phi_{01} - \rho'\phi_{11} -{\textstyle\frac32}\eth\psi\notag\\
 &=\Phi_{12}\Hl+\Phi_{01}\Hn -\left(\Phi_{11}+3\Pi\right)\Hm -\Phi_{02}\Hmb
&\tag{\themaxintcons$b$}\\[1pt]
\eth'\phi_{01} +\thorn\phi_{00} &- \rho\phi_{00} -
  2\tau'\phi_{01} - \sigma'\phi_{11}+{\textstyle\frac32}\eth'\psi\notag\\
 &= -\Phi_{21}\Hl-\Phi_{10}\Hn+\Phi_{20}\Hm+\left(\Phi_{11}+3\Pi\right)\Hmb
&\tag{\themaxintcons$c$}\\[1pt]
\thorn'\phi_{01}+\eth\phi_{00}& - \tau\phi_{00} -
  2\rho'\phi_{01} -\kappa'\phi_{11}+{\textstyle\frac32}\thorn'\psi\notag\\
& =-\Phi_{22}\Hl -\left(\Phi_{11}-3\Pi\right)\Hn+\Phi_{21}\Hm+\Phi_{12}\Hmb
\tag{\themaxintcons$d$}
\end{align}

All these equations are easily checked to be consistent as far as spin and boost
weight are concerned, and reduce to the equations of [14] for a homothety in
vacuum.

\section{Second integrability conditions}

Since a conformal transformation preserves the conformal structure, 
${\mathcal L}_\xi C^a{}_{bcd}=0$, and resolving the spinor version of this equation
%and using ($\theintcons$) to eliminate first derivatives of the $\phi_{ij}$
leads to second integrability conditions involving the $\Psi_i$. 
The Ricci tensor is not preserved under a conformal transformation, but 
instead we have, cf.~[6], for the trace-free Ricci tensor and Ricci scalar
\begin{equation}
  {\mathcal L}_\xi S_{ab} = 2\psi_{;ab}- \frac12(\psi_{;cd}g^{cd})g_{ab}\qquad
 {\mathcal L}_\xi R = -2\psi R + 6 (\psi_{;cd}g^{cd}) .\label{eq:RicciConf}
\end{equation}

Resolving these gives second integrability conditions involving
the $\Phi_{ij}$ and $\Pi$.
The same integrability conditions arises from applying
the commutators to the components of the conformal bivector of course. 
Using the Bianchi identities and the GHP notation these equations can be reduced to a
compact form. Firstly, define the zero weight derivative operator
$$\GHPL_\xi= \Hn\thorn+\Hl\thorn'-\Hm\eth'-\Hmb\eth,$$
and let
$$X_{00}=\phi_{00}-\kappa'\Hl-\tau'\Hn+\sigma'\Hm+\rho'\Hmb,\qquad
X_{11}=\phi_{11}+\kappa\Hn+\tau\Hl-\sigma\Hmb-\rho\Hm.$$
(Note that under the Sachs $*$ operation, $X_{11}$ and $X_{00}$ are unchanged
but $\bar{X}_{11}^*=\bar{X}_{00}$ and $\bar{X}_{00}^*=\bar{X}_{11}$).
Then we find that for the Weyl tensor components
\begin{equation}
  \GHPL_\xi \Psi_i + 2\psi\Psi_i = iX_{00}\Psi_{i-1} -2(2-i)\,\phi_{01}\Psi_i +
(i-4)X_{11}\Psi_{i+1}\label{eq:Psitwo},
\end{equation}
The Ricci tensor components are more involved because of the presence of the second
derivatives of the divergence. I will write them as 
\newcounter{Phitwo}
\stepcounter{equation}
\setcounter{Phitwo}{\value{equation}}
\begin{align}
\GHPL_\xi \Phi_{ab} + 2\psi\Phi_{ab}+ \Upsilon_{ab}\psi &= 
   aX_{00}\Phi_{(a-1)b} + b\bar X_{00}\Phi_{a(b-1)} %\notag \\
{}+ (a-2)X_{11}\Phi_{(a+1)b}  \notag\\
 &\quad{}+ (b-2)\bar X_{11}\Phi_{a(b+1)} - 2((1-a)\,\phi_{01} +
(1-b)\,\bar\phi_{01})\Phi_{ab} \tag{\thePhitwo} \\
\GHPL_\xi\Pi +2\psi\Pi & =\frac12\left(\thorn'\thorn-\eth'\eth
-\bar\rho'\thorn-\rho\thorn' +\bar\tau\eth + \tau\eth'\right)\psi
 ,\label{eq:Pitwo}
\end{align}
where $\Upsilon_{ab}$ are differential operators given below.
%Note that $\Psi_i$ has PR-weight $[3-i,i-1,1,1]$ and $\Phi_{ab}$ weight
%$[2-a,a,2-b,b]$.
Equations~(\ref{eq:Psitwo}) are equivalent to Collinson and French's equations
($2.2$) [3] and Kolassis and Ludwig's equations (43)--(45) [9]; 
equations~(\thePhitwo) are equivalent to [9] equations (47)--(49). 

The operators $\Upsilon_{ab}$ have the same symmetry properties under conjugation as
$\Phi_{ab}$ and are
\begin{align*}
  \Upsilon_{00} &= \thorn^2+\bar\kappa\eth+\kappa\eth' \notag\\
\Upsilon_{01} & = \eth\thorn+\bar\rho\eth+\sigma\eth' =
\thorn\eth +\tau'\thorn+\kappa\thorn' \\
 \Upsilon_{02} & = \eth^2 + \bar\sigma'\thorn + \sigma\thorn'\\
\Upsilon_{11} & = \frac12\left(\thorn'\thorn+\eth'\eth
+\bar\rho'\thorn+\rho\thorn'+\bar\tau\eth + \tau\eth' \right) \\
\Upsilon_{12} & = \eth\thorn'+\rho'\eth+\bar\rho'\eth'=
\thorn'\eth +\bar\kappa'\thorn+\tau\thorn' \\
\Upsilon_{22} & = (\thorn')^2 +\kappa'\eth+\bar\kappa'\eth'
\end{align*}
along with the complex conjugates. The alternate forms here arise from the
commutators.

As Geroch points out in the appendix to [4], see also [6], a conformal vector is
given locally by its values and first two derivatives at a point. So there are 
no further true integrability conditions.

\section{Surface homogenous metrics}

Brinkmann's Theorem (see e.g.~[16]), which follows from
equation~(\ref{eq:RicciConf}), proves the only vacuum metrics
with a proper conformals are pp~waves, which have been much studied, see
e.g.~[12]. So for an application  we consider the case of a metrics with a $G_3$ of
motions on a spacelike surface [16], compare [17] and references therein which
classified all the spherical symmetric cases. We will begin with the metric
in null
coordinates:
\begin{equation}
  ds^2 = 2e^{2F(u,v)}du\,dv -
e^{2X(u,v)}\left(dx^2+\Sigma^2(x)dy^2\right).\label{eq:surfhom}
\end{equation}
Here $\Sigma(x)=\sin x$ for the spherically symmetric case, $\Sigma(x)=\sinh(x)$ for
pseudo-spherical case and $\Sigma(x)=1$ for the plane symmetric case. The
(isometric) isotropy implies such metrics are Petrov type D or O and the
Ricci tensor has at least two equal eigenvalues which must correspond to
spacelike eigenvectors [16]. The
Kimura metric considered as an example in [11]
is a special case of this metric (see later).

The obvious normalised Newman-Penrose tetrad
$$\ell_a= e^F\,dv,\quad
 n_a= e^F\,du,\quad
 m_a = \frac1{\sqrt2}e^X\left(dx+i\Sigma(x)\,dy\right),\quad
$$
is a Petrov canonical tetrad: of the $\Psi_i$ only 
$$\Psi_2  = -\frac13e^{-2F}\left(X_{,uv} - F_{,uv}\right) +
\frac16e^{-2X}\Sigma_{,xx}\Sigma^{-1}$$ 
is non-zero and is also
close to a Ricci canonical tetrad, as only $\Phi_{00}$,
$\Phi_{11}$, $\Phi_{22}$ and $\Pi$ are not identically zero (see appendix) so the
Ricci tensor is
diagonalisable, but not necessarily diagonalisable over $\R$ .
We find that the only non-zero spin coefficients of good weight are
$$ \rho = e^{-F}X_u ,\qquad
  \rho' = -e^{-F}X_v ,
$$
with the other spin coefficients also being real.

As $\Sigma_{xx}/\Sigma=-\frac12K$ where $K$ is the (constant)
Gaussian curvature of the Killing orbits, we can see that these metrics
are conformally flat iff $X-F = \log|u-Kv|$ for $K=\pm1$, or separable
($X-F=P(u)+Q(v)$) for $K=0$.
We assume from hence that the metric is type D, since the conformally flat case is
known to have 15 independent conformal vectors. See also [2] for an analysis of
these cases.

Suppose we have a conformal vector in a type D surface homogenous
space-time~(\ref{eq:surfhom}). Then
equations~(\ref{eq:Psitwo}) for $i=1,3$ give us $\phi_{11}=\rho\Hm$ and
$\phi_{00}=-\rho'\Hmb$. If we substitute these expressions into equation
(\theintcons$d$) and use the conformal equations and
the curvature equation $\thorn'\rho=\rho\rho'-\Psi_2-2\Pi$, we find that
$\eth\psi=0$, and $\eth'\psi=0$ follows as $\psi$ is real and
zero-weighted.

We can recast (\theintcons$c$) too: we find that 
$$\thorn\psi = \Hl\left(-\rho\rho'+\Psi_2+2\Pi\right)- 
\Hn\left(\rho^2+\Phi_{00}\right) -\rho\left(\phi_{01}+\bar\phi_{01} -\psi\right).$$
But the conformal equations and the curvature equations mean that this is equivalent
to
$$ \thorn\left( \psi + \rho'\Hl+\rho\Hn\right)=0.$$
Similarly, (\theintcons$f$) gives $ \thorn'\left( \psi + \rho'\Hl+\rho\Hn\right)=0$.
But each term in the real zero-weighted scalar ${\cal C}=\psi + \rho'\Hl+\rho\Hn$ is
also annihilated by $\eth$ and $\eth'$, and hence $\cal C$ is a constant.

Using equations (\theintcons$i$) to (\theintcons$l$) we can now find explicit
equations for the derivatives of $\phi_{01}$. If we set $\phi_{01} = A+iB$ for real
weight (0,0) scalars $A$ and $B$ we get
$$\thorn B=\thorn'B=\eth A=\eth'A=0$$
and 
\begin{align}
  \eth B & = -i\Hm\left(\rho\rho'+\Psi_2-\Phi_{11}-\Pi\right)\label{eq:ethB} \\
\thorn A & = \frac12\Bigl(\Hl(3\Psi_2+2\Phi_{11})-\Hn\Phi_{00}-
           \rho(2A+{\cal C})\,\Bigr)\label{eq:thA}\\
\thorn' A & = \frac12\Bigl(-\Hn(3\Psi_2+2\Phi_{11})+\Hl\Phi_{22}-
           \rho'(2A-{\cal C})\Bigr) \label{eq:thpA}.
\end{align}
The conformal equations now take the form
\begin{align}
 \thorn'\Hl&=-2A-\rho'\Hl-\rho\Hn+\cal C,\label{eq:thpKl}\\
\thorn\Hn & = 2A-\rho'\Hl-\rho\Hn+\cal C,\label{eq:thKn}\\
\thorn\Hm &=-\rho\Hm, \qquad \thorn'\Hm=-\rho'\Hm\label{eq:thKm}\\
\eth'\Hm &= 2iB-\cal C, \label{eq:epHm} \\ 
\thorn\Hl&=\eth\Hl =\thorn'\Hn=\eth\Hn=\eth\Hm=0\label{eq:ethall},
\end{align}
plus their conjugates.

The only equations left to be considered are the remaining second integrability
equations. Most of these turn out to be already satisfied modulo the conformal
equations, curvature equations and Bianchi identities. If we also make use of the
commutators and the fact that $\eth$ and $\eth'$ annihilate all the spin coefficients
and curvature components --- which is most easily checked from the coordinate form of
the metric --- we find there is only one second integrability
equation left:
\begin{equation}
 % (\thorn'\Psi_2-2\rho'\Psi_2)\Hl + (\thorn\Psi_2-\rho\Psi_2)\Hn 
%\eth\Psi_2\,\Hmb+\eth'\Psi_2\,\Hm 
%&= -2\Psi_2{\cal C}\label{eq:CCON} \\ 
4\bigl(\rho\rho' + \Psi_2-\Pi -\Phi_{11}\bigr){\cal C }=0. \label{eq:Xnalg}
\end{equation}

Now the term in brackets on the right hand side of~(\ref{eq:ethB}) and the 
left hand side of~(\ref{eq:Xnalg})
occurs in the commutators, so can be written in terms of spin coefficients not of
good weight: it is
$4\alpha^2-2\delta\alpha=\frac12e^{-2X}\Sigma_{xx}/\Sigma=
-\frac12Ke^{-2X}$, where $K$ is the (constant) Gaussian curvature of the 
Killing orbits [15]. 
So in the plane symmetric case we have $B$ constant from~(\ref{eq:ethB}) etc, and
in the other cases ${\cal C}=0$ from~(\ref{eq:Xnalg}).

\medskip
The 6 unknowns fall naturally into disjoint sets $\{\Hl,\Hn,A\}$ and
$\{\Hm,\Hmb,B\}$, where everything in the
first set is annihilated by both $\eth$ and $\eth'$, which is easily seen to be
equivalent to being independent of $x$ and $y$.

In the non-plane cases, or wherever ${\cal C}=0$, these sets are completely
decoupled.
But it is a simple matter to solve for the terms $\Hm$ and $\Hmb$ in
the plane symmetric case to get (with $\Sigma(x)=1$, i.e.~Cartesian coordinates)
$$-\Hm \mb^a -\Hmb m^a= 2B(y\partial_x-x\partial_y) +k_1\partial_x+k_2\partial_y+
{\cal C}(x\partial_x+y\partial_y),
$$
where constants $B$, $k_1$ and $k_2$ give the known Killing vectors in the spacelike
surface. These are the only possible conformals vectors lying completely in
the Killing orbits, as $\Hl=\Hn=0$ implies $A={\cal C}=0$. The term with ${\cal C}$
is the standard homothety in the plane, which from the equations is coupled to the
components orthogonal to the Killing orbit. That such coupling must be present also
follows from the fact that
we cannot have a conformal vector
with a fixed point at which the divergence in non-zero in Petrov type D [7]: as
${\cal C}=\psi + \rho'\Hl+\rho\Hn$, any conformal vector with ${\cal C}$ non-zero
cannot be tangent to the orbit of the known Killing vector.
Since
we can subtract Killing vectors from any proper conformal vector, it follows
that in looking for other
conformal vectors in a plane symmetric metric we can assume we have $B=0$, 
$\Hm={\cal C}e^{X}(x+iy)/\sqrt2$ and solve the remaining equations for $\Hl$, $\Hn$
and $A$.

\medskip
For the non-plane cases
%the set $\{\Hm,\Hmb,B\}$,
the left hand side of equation~(\ref{eq:ethB}) is 
$\frac12 iKe^{-2X}\Hm$, where the Gaussian curvature $K=\pm1$. So using
equation~(\ref{eq:epHm})
$\eth'\eth B = - Ke^{-2X}B$, or in coordinates $\nabla^2B=-2KB$ for
$\nabla^2$ the Laplacian on the Killing orbit. As is well-known this equation
has exactly three independent solutions.
%arising from the three possible homogeneous degree 1 polynomials in $\R^3$. 
In our coordinates they are
$$ \Sigma'(x),\quad \Sigma(x)\sin y,\quad\text{and}\quad \Sigma(x) \cos y.$$
We have now solved the $\{\Hm,\Hmb,B\}$ set, since $\Hm=2ie^{2X}\eth B$:
the only conformal vectors with $\Hn=\Hl=0$ are again the 
three known Killing vectors. Similarly to the plane symmetric case, we can restrict
the search for proper conformal vectors in the non-plane cases to those
vectors with $\Hm=B=0$.

% Before converting to coordinate form, note that $\psi={\cal
% C}-(\rho'\Hl+\rho\Hn)$and that at a fixed point of a conformal vector in type D the
% divergence must vanish [7].

At this stage it is as well to convert to a coordinate form of the main conformal
equations. Firstly, equations~(\ref{eq:ethall}) implies that 
$$\Hl = Y^2(v)e^{F(u,v)}\quad \text{and}\quad 
 \Hn =  Y^1(u)e^{F(u,v)}.$$
The surface orthogonal part of the conformal vector, $\Hl n^a+\Hn\ell^a$, is
then $Y^1(u)\partial_u + Y^2(v)\partial_v$ and $\psi ={\cal C}+ Y^1X_{,u} +
Y^2X_{,v}={\cal C}+\xi^aX_{,a}$.

The other derivatives of $\Hn$ and $\Hl$ can be combined to give expressions for $A$
and $\cal C$. Writing $Z$ for $F-X$ equations~(\ref{eq:thpKl}) and (\ref{eq:thKn})
give
\begin{align}
  4A &= Y^1_{,u} +2Y^1 F_{,u} - Y^2_{,v} - 2Y^2F_{,v} 
 %= e^{-2F}\Bigl( (e^{2F}Y^1)_{,u} - (e^{2F}Y^2)_{,v}\Bigr)
\label{eq:AZ} \\
  2{\cal C} &= Y^1_{,u} + 2Y^1Z_{,u} + Y^2_{,v} + 2Y^2Z_{,v}
 % = e^{2Z}\Bigl( (e^{-2Z}Y^1)_{,u} +  (e^{-2Z}Y^2)_{,v}\Bigr) 
\label{eq:CeqZ}
%  2{\cal C} &= Y^1_{,u} +2Y^1 (F_{,u}-X_{,u}) + Y^2_{,v} + 2Y^2(F_{,v}-X_{,v})\\
% & = e^{2(X-F)}\Bigl( (e^{2(F-X)}Y^1)_{,u} +  (e^{2(F-X)}Y^2)_{,v}\Bigr)
\end{align}
These expressions for $A$ and $\cal C$ can then be substituted into the
integrability conditions~(\ref{eq:thA}) and
(\ref{eq:thpA}) to give
\begin{align}
  Y^1_{,uu} + 2Z_{,u}Y^1_{,u}+ 2Z_{,uu}Y^1 &=
-2Z_{,uv}Y^2 \label{eq:Q5} \\
  Y^2_{,vv} + 2Z_{,v}Y^2_{,v} + 2Z_{,vv}Y^2 &=
-2Z_{,uv}Y^1. \label{eq:Q6}
\end{align}
However, in coordinate form it is easy to see that both these latter equations
are identically satisfied modulo equation~(\ref{eq:CeqZ}), a fact that can be proved
in the GHP notation with sufficient work. Since we can consider
equation~(\ref{eq:AZ}) as defining $A$, with a little algebra we have, cf.~[13]:

\begin{theorem}
  The surface homogeneous metric
\begin{equation}
  ds^2 = e^{2X(u,v)}\left(2e^{2Z(u,v)}du\,dv -
\bigl(dx^2+\Sigma^2(x)dy^2\bigr)\,\right),\label{eq:spaceC}
\end{equation}
if type D, admits the conformal vector 
$$ \xi^a = Y^1(u)\partial_u+ Y^2(v)\partial_v + 
 {\cal C}\left(x\partial_x+y\partial_y\right)$$
%with divergence ${\cal C}+(F+Z)_{,a}\xi^a$
iff the equation
\begin{equation}
2{\cal C} = Y^1_{,u} + 2Y^1Z_{,u} + Y^2_{,v} + 2Y^2Z_{,v}
 % = e^{-2Z}\Bigl( (e^{2Z}Y^1)_{,u} +  (e^{2Z}Y^2)_{,v}\Bigr)
\tag{\ref{eq:CeqZ}}
\end{equation}
is satisfied for constant $\cal C$, which must be zero in the non-plane symmetric
cases. 

The divergence $\psi$ of $\xi^a$ is then
${\cal C}+\xi^aX_{,a}=(X+Z)_{,a}\xi^a+\frac12\left(Y^1_{,u}+Y^2_{,v}\right)$.
\end{theorem}

The form of the metric in this theorem is doubly convenient in that it shows these
metrics to be conformally reducible, in the sense of Carot and Tupper [2], by
illustrating the conformal scaling which is irrelevant to the conformal
vector: we expect a free conformal factor in the metric if all we do is look for
conformal vectors. This factor would be fixed by the field equations. 

Tupper {\sl et al} [17] classified the spherically symmetric cases according to the
possible extra Killing or homothetic vectors. Their analysis applies with no 
change to the case where the Killing orbit has negative curvature. 

Here I will merely point out that by the Bilyalov--Defrise-Carter Theorem [1,7]
and isotropy considerations, the maximal size of the conformal algebra is 6, and in
these cases the metric is locally conformal to a metric with a
6-parameter group of
motions, and hence locally conformal to a product of two 2-spaces of constant
curvature, [16]. Any metric~(\ref{eq:spaceC}) with $Z_{uv}=0$, for example, is
easily shown to have a 6 parameter conformal algebra. 

%The form of the conformal vector given here is vastly simpler than given in [13],
%for example, but of course the physical interpretation of the metric is much harder
%in the null coordinates. If the metric were to be a fluid, that is of Segre type
%$\{1,111\}$, with the fluid flow vector a timelike eigenvector of the
%Ricci tensor, that vector would be
%$$ u^a = \frac1{\sqrt2}\left(k\ell^a+k^{-1}n^a\right)\quad\text{where}\quad
%k^4=\Phi_{22}/\Phi_{00}.$$
 
One explicit example with the maximal conformal algebra is the spherically symmetric
Kimura case [8] considered in [10,11]. Changing the
coordinates in those references to $u =t - r^{-1}$, $v=t+r^{-1}$ we have
$F=-\log(u-v)+\log(b_0)$, $Z=-\log b_0$ in~(\ref{eq:surfhom}) and
solving~(\ref{eq:CeqZ}) with ${\cal C}=0$ we have
$$ Y^1(u)= k_1u+k_2+k_3,\qquad Y^2(v)=-k_1v-k_2+k_3,$$
which are easily seen to give the two proper conformal vectors ($k_1$ and $k_2$) and
fourth Killing vector ($k_3$) given by in [10,11]. 

\bigskip
A similar calculation can be performed in the case of a metric with a
three-parameter Killing group on time-like surfaces. We start with the metric in the
form
\begin{equation}\label{eq:homT2}
  ds^2 = e^{2X(\zeta,\bar\zeta)}\left(\Sigma^2(r)dt^2-dr^2\right) -
 2e^{2F(\zeta,\bar\zeta)}d\zeta\,d\bar\zeta
\end{equation}
see [16], but note I have chosen conformally flat coordinates in the
surface orthogonal to the Killing orbit. Since the metrics in the two cases are the
same under the complex coordinate transformations 
$$ (u,v,x,y) \mapsto (\zeta,\bar\zeta,ir,it)$$
plus a switch of signature (essentially the Sachs $*$ operation), similar results
arise.

The obvious normalised tetrad
\begin{equation*}
\ell_a=\frac1{\sqrt{2}}\,e^{X}\left(\Sigma  \,dt+dr\right),\quad
n_a=\frac1{\sqrt{2}}e^{X}\left(\Sigma  \,dt-dr\right) ,\quad
m_a=e^{F}d\zeta
\end{equation*}
is again a Petrov canonical tetrad, with only $\tau=\bar\tau'$ non-zero spin
coefficients and $\Psi_2$, $\Phi_{11}$, $\Phi_{02}$,
$\Phi_{20}$, $\Pi$ non-zero in the curvature. The unknowns split into the same two
disjoint sets $\{\Hl,\Hn,A\}$ and $\{\Hm,\Hmb,B\}$ where $\phi_{01}=A+iB$. 
with everything in the second set annihilated by $\thorn$ and $\thorn'$. 
The
equations for the first set solve to give the three Killing vectors, and so only the
equations for the second set remain:
\begin{align*}
\thorn\Hm&=\thorn'\Hm=\eth\Hm=0\\
\eth'\Hm&=2iB-\tau\Hmb-\bar\tau\Hm- {\cal C}\\
2\eth B&=-2\tau B-i\Phi_{02}\Hmb-i(3\Psi_2-2\Phi_{11})\Hm
\end{align*}
plus their conjugates, for constant $\cal C$ which is zero if the Killing orbit
is not flat. All other equations are identically satisfied modulo these
and the curvature equations etc as before.

Writing the spatial part of the conformal vector as
$-\Hm\mb^a-\Hmb m^a = Y\partial_\zeta+\bar Y\partial_{\bar\zeta}$, the $\thorn$ and
$\thorn'$ equations show that $Y$ is a function of $(\zeta,\bar\zeta)$
only, and the remaining equations reduce to the Cauchy-Riemann equations,
$ Y_{,\bar\zeta} = 0$
and the equivalent of equation~(\ref{eq:CeqZ})
\begin{equation}\label{eq:CeqZT2}
2{\cal C}= Y_{,\zeta}+\bar{Y}_{,\bar\zeta} +2(YZ_{,\zeta} + 2\bar{Y}Z_{,\bar\zeta}),
\end{equation}
where $Z(\zeta,\bar\zeta)=X(\zeta,\bar\zeta)-\lambda(\zeta,\bar\zeta)$, or serve to
define $B$:
$$ 
 4B(\zeta,\bar\zeta)=i\left(
  Y_{,\zeta}-\bar{Y}_{,\bar\zeta}+2Y
\lambda_{,\zeta}-2\bar{Y}\lambda_{,\bar\zeta}\right).
$$
The appearance of the Cauchy-Riemann equations is to be expected, since we are
essentially looking for conformals on a Riemannian 2-space. So

\begin{theorem}
  The timelike-surface homogeneous metric
\begin{equation*}
  ds^2 = e^{2X(\zeta,\bar\zeta)}\left(
  \Sigma^2(r)dt^2-dr^2 -
 2e^{2Z(\zeta,\bar\zeta)}d\zeta\,d\bar\zeta
 \,\right),
\end{equation*}

admits the conformal vector 
$$ \xi^a = Y(\zeta)\partial_\zeta+ \bar{Y}(\bar\zeta)\partial_{\bar\zeta} + 
 {\cal C}\left(t\partial_t+r\partial_r\right)$$
%with divergence ${\cal C}+(F+Z)_{,a}\xi^a$
iff $Y(\zeta)$ is analytic and the equation
\begin{equation}
2{\cal C}= Y_{,\zeta}+\bar{Y}_{,\bar\zeta} +2(YZ_{,\zeta} + 2\bar{Y}Z_{,\bar\zeta})
\tag{\ref{eq:CeqZT2}}
\end{equation}
is satisfied for constant $\cal C$, which must be zero in the non-plane symmetric
cases. 

The divergence $\psi$ of $\xi^a$ is then
${\cal
C}+\xi^aX_{,a}=(X+Z)_{,a}\xi^a+\frac12\left(Y_{,\zeta}+\bar{Y}_{,\bar\zeta}\right)$.
\end{theorem}

An analysis similar to that of [17] for the spherically symmetric situation can
obviously be carried out in this case too.

As in the spherical case we can  find a type D metric of the
form~(\ref{eq:homT2}) 
admitting the maximal 6-parameter conformal group. One such example is given by
setting $X(\zeta,\bar\zeta)=\ln(\zeta+\bar\zeta)+\lambda(\zeta,\bar\zeta)$ when the
metric admits, in addition to the Killing vectors, the conformal vectors
$$\xi^a = \left(k_1i\zeta^2+k_2\zeta+k_3i\right)\partial_\zeta +
% \left(-k_1i\bar\zeta^2+k_2\bar\zeta-k_3i\right)\partial_{\bar\zeta}
\text{conjugate}
$$
with $k_i$ real constants.

\section*{Appendix: Components of the Curvature}

In the surface-homogeneous metric~(\ref{eq:surfhom}) with Newman-Penrose tetrad
as given the non-zero Ricci tensor components are
\begin{align*} 
 \Phi_{11} & = \frac12e^{-2F}\left(X_{,u}X_{,v} - F_{,uv} \right)
 -\frac14 e^{-2X}\Sigma_{,xx}\Sigma^{-1},
 \\
 \Phi_{00}=\Phi_{22} & = e^{-2F}\left(2 F_{,u}X_{,u} - X_{,uu} - X_{,u}{}^2
 \right),\\
 \Pi & =\frac16e^{-2F}\left( F_{,uv} + 2X_{,uv} + 3 X_{,u}X_{,v} \right)
  -\frac1{12}e^{-2X}\Sigma_{,xx}\Sigma^{-1}.
\end{align*}

In the timelike case, metric~(\ref{eq:homT2}), with Newman-Penrose tetrad
as given the non-zero curvature components are
\begin{align*} 
 \Psi_2 &= \frac13e^{-2\lambda}\left(\lambda-X\right)_{,\zeta\bar\zeta} +
 \frac16 e^{-2X}\Sigma_{,rr}\Sigma^{-1},\\
 \Phi_{11} & = \frac12e^{-2\lambda}
\left(X_{,\zeta}X_{,\bar\zeta}-\lambda_{,\zeta\bar\zeta} \right) + 
 \frac14 e^{-2X}\Sigma_{,rr}\Sigma^{-1},\\
 \Phi_{20}=\bar{\Phi_{02}} & = e^{-2\lambda}\left( 2 X_{,\zeta}\lambda_{,\zeta} -
  X_{,\zeta}{}^2 - X_{,\zeta\zeta} \right), \\
  \Pi & = -\frac16e^{-2\lambda}\left( \lambda_{,\zeta\bar\zeta} + 
  2X_{,\zeta\bar\zeta} + 3 X_{,\zeta}X_{,\bar\zeta}  \right) -
  \frac1{12} e^{-2X}\Sigma_{,rr}\Sigma^{-1}.
\end{align*}

\section*{Acknowledgements}

Calculations were carried out using Maple, and in particular the GHPII package
of Vu and Carminati [18]. Maple is a registered trademark of Waterloo Maple Inc.

\section{References}

\begin{enumerate}
\item[{[1]}] Bilyalov RF {\it Sov.~Phys.~Dok} {\bf 8} (1964) p.~878; Defrise-Carter
L {\it Comm.~Math.~Phys.} {\bf 40} (1975) p.~273
\item[{[2]}] Carot J and Tupper BOJ, {\it Class.~Quant.~Phys.} {\bf 19} (2002)
p.~4141 
\item[{[3]}] Collinson CD and French DC {\it J.~Math.~Phys.} {\bf 8} (1967)
p.~701
%\item[{[EL]}] Edgar SB and Ludwig G {\it Gen.~Rel.~Grav.} {\bf 29} (1997) 
%p.~1309
\item[{[4]}] Geroch R {\it Comm.~Math.~Phys} {\bf 13} (1969) p.~180
\item[{[5]}] Geroch R, Held A and Penrose R {\it J.~Math.~Phys.} {\bf 14}  (1973)
p.~874
\item[{[6]}] Hall GS (2004) {\it Symmetries and Curvature Structure in General
Relativity} (Singapore: World Scientific Publishing)
\item[{[7]}] Hall GS and Steele JD {\it J.~Math.~Phys} {\bf 32} (1991) p.~1847
\item[{[8]}] Kimura M {\it Tensor} {\bf 30} (1975) p.~395
\item[{[9]}] Kollasis C and Ludwig G {\it Gen.~Rel.~Grav.} {\bf 25} (1993) p.~625
\item[{[10]}] Koutras A {\it Class.~Quantum Grav.} {\bf 9} (1992) p.~1573
\item[{[11]}] Ludwig G {\it Class.~Quantum Grav.} {\bf 19} (2002) p.~3799
\item[{[12]}] Maartens and Maharaj SD {\it Class.~Quant.~Grav.} {\bf 8} (1991)
p.~503
\item[{[13]}] Moopanar S and Maharaj SD {\it Int.~J.~Theor.~Phys} {\bf 49} (2010)
p.~1878
\item[{[14]}] Steele JD  {\it Class.~Quantum Grav.} {\bf 19} (2002) p.~259
\item[{[15]}] Penrose R and Rindler W 1984 {\it Spinors and Space-Time} vol 1
(Cambridge: Cambridge University Press)
\item[{[16]}] Stephani H, Kramer D, MacCallum M, Hoenselaers C and Herlt E
2003
{\it Exact Solutions to Einstein's Field Equations, 2nd Edition} (Cambridge:
Cambridge University Press)
\item[{[17]}] Tupper BOJ, Keane A and Carot J {\it Class Quant Grav} {\bf 29}
(2012) 145016
\item[{[18]}] Vu K and Carminati J {\it Gen.~Rel.~Grav.} {\bf 35} (2003) p.~263
\end{enumerate} 

\end{document}